\documentstyle[prl,aps,epsf,multicol]{revtex}
\begin{document}
\tighten
\title{Simple Microscopic Theory of Amontons' Laws for Static Friction}
\author{M. H. M\"user$^1$, L. Wenning$^1$, and M. O. Robbins$^2$}

\address{$^1$ Institut f\"ur Physik, WA 331; 
              Johannes Gutenberg Universit\"at;
              55099 Mainz; Germany \\
         $^2$ Dept. of Physics \& Astronomy;
              Johns Hopkins University;
              Baltimore, MD 21218; USA \\
        }

\date{\today}
\maketitle

\begin{abstract}   
A microscopic theory for the ubiquitous phenomenon of static friction
is presented.
Interactions between two surfaces are modeled by an energy penalty
that increases exponentially with the degree of surface overlap.
The resulting static friction is proportional to load, in accordance
with Amontons' laws.
However the friction coefficient between bare surfaces vanishes
as the area of individual contacts grows,
except in the rare case of commensurate surfaces.
An area independent friction coefficient is obtained for any surface
geometry when an adsorbed
layer of mobile atoms is introduced between the surfaces.
The predictions from our simple analytic model are confirmed by atomistically 
detailed molecular dynamics simulations.
\end{abstract}

\begin{multicols}{2}
\narrowtext

Static friction $F_{\rm s}$ is the lateral force that must be applied
to initiate sliding of one object over another.
Its presence implies that the objects have locked together into a local
energy minimum that must be overcome by the external force.
This threshold force is observed between all the objects around us,
yet its molecular origins have remained baffling.
How do surfaces manage to lock together, and why does the friction
obey Amontons' laws that
$F_{\rm s}$ increases linearly with the load $L$ pushing the objects
together, and is independent of the
surface area
\cite{bowden86}?

\begin{figure}[hbtp]
\begin{center}
\leavevmode
\hbox{ \epsfxsize=50mm \epsfbox{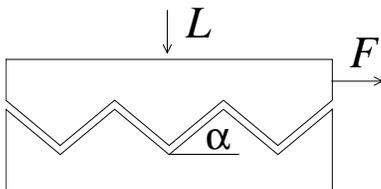}}
\caption{Sketch of two surfaces with interlocking asperities.
The top surface experiences
a normal load $L$ and a lateral force $F$, which attempts to pull the top
surface up the slope $\tan\alpha$. The bottom wall is fixed.
}
\label{fig:slope}
\end{center}
\end{figure}

Early theories of friction were based on the purely geometric
argument that friction is caused by interlocking of
surface asperities~\cite{bowden86,HISTORIC}.
The idea (Fig. \ref{fig:slope}) is that
the top surface must be lifted up a typical slope $\tan \alpha$
determined by roughness on the bottom surface.
If there is no microscopic friction between the surfaces, then
the minimum force to initiate sliding is
$F_{\rm s} = L \tan \alpha $.
This result satisfies Amontons' laws with a constant coefficient
of friction $\mu_{\rm s} \equiv F_{\rm s}/L = \tan \alpha$.
In 1737,
B{\'e}lidor obtained a typical experimental value of
$\mu_{\rm s} \approx 0.35$ by modeling rough surfaces as spherical
asperities arranged to form commensurate crystalline walls \cite{HISTORIC}.
More recently, Israelachvili {\it et al.} have discussed a
similar ``cobblestone'' model where the spherical asperities are
atoms\cite{israel}.
However, asperities on real surfaces
do not match as well as envisioned in these models
or sketched in Fig. \ref{fig:slope}.
On average,
for every asperity or atom going up a ramp, there is another going down.
One concludes that the mean friction between rigid surfaces vanishes
unless they happen to have the same periodicity and alignment.
Detailed calculations show that elastic deformations are generally
too small to alter this
conclusion \cite{hirano90,sorensen96,muser00,robbins00}.

In this Letter, we study a simple model of the static friction in individual
contacts as a function of their area $A$.
At fixed $A$,
the static friction between bare surfaces is proportional to load.
However, as implied by
previous work \cite{hirano90,sorensen96,muser00,robbins00},
the prefactor vanishes for incommensurate surfaces.
We find that $F_{\rm s}/L$ also vanishes for disordered surfaces,
scaling as $A^{-1/2}$ as the area increases.
These results for bare surfaces are inconsistent with Amontons' laws.
Of course most surfaces around us are covered by a few angstroms of
hydrocarbons, water and other small airborn molecules \cite{he99}.
We show that
introducing such mobile molecules into the interface
between incommensurate or disordered surfaces
yields a value of $\mu_{\rm s}=F_{\rm s}/L$ that is independent of contact size
and load.
This naturally leads to Amontons' laws for any distribution
of contacts.
Our analytic predictions are then tested against
previous \cite{he99} and new computer simulations.

In our model, two surfaces ``pay'' an energy penalty $V_{\rm ww}$
that increases exponentially as the local separation between them
decreases and they begin to overlap.
This form of interaction is frequently used to model the Pauli
repulsion between atoms, and contains the hard-sphere
models of B{\'e}lidor and Israelachvili {\it et al.} as a limiting case.
Within the contact area $A$
the surfaces are parallel to the $xy$ plane.
The bottom wall is held fixed, has zero mean height, and local
height $\delta z_{\rm b}(\vec{x})$.
The mean height $z_{\rm t}$ and lateral position $\vec{x}_{\rm t}$
of the top wall are varied, and its local height is
$z_{\rm t}+\delta z_{\rm t}(\vec{x}-\vec{x}_{\rm t})$.
Then $V_{\rm ww}$ can be written as an integral over $A$
\begin{equation}
\label{eq:penalty}
V_{\rm ww}(\vec{x}_{\rm t},z_{\rm t}) = \varepsilon
\int d^2x \;
e^{-\left[z_{\rm t} + \delta z_{\rm t}(\vec{x}-\vec{x}_{\rm t})-
\delta z_{\rm b}(\vec{x})
          \right]/\xi},
\label{eq:def_ene}
\end{equation}
where $\xi$ and $\varepsilon$ characterize the length and energy
per unit area of the interaction.
The dependence of $V_{\rm ww}$ on $z_{\rm t}$ can be pulled out
of the integral in Eq. \ref{eq:def_ene}.
The integral then depends only on $\vec{x}_{\rm t}$
and can be expressed as an effective shift in the mean wall separation by
$\Delta z(\vec{x}_{\rm t})$:
\begin{equation}
V_{\rm ww}(\vec{x}_{\rm t},z_{\rm t})
= \varepsilon A \; e^{ 
 -\left\{ z_{\rm t} + \Delta z(\vec{x}_{\rm t}) \right\}/ \xi} . 
\label{eq:eq2}
\end{equation}

The lateral and normal forces on a static contact are equal to the derivatives
of $V_{\rm ww}$ with respect to
$\vec{x}_{\rm t}$ and $z_{\rm t}$, respectively.
Due to the simple form of Eq. \ref{eq:eq2} almost all factors
in the ratio of friction to normal load cancel, and the ratio
is independent of load \cite{volmer97}.
The static friction coefficient corresponds to the maximum of
this ratio, taken along the direction of the applied force:
\begin{equation}
\mu_{\rm s} = \max \left[ {\partial \over \partial x_{\rm t} }
\Delta z(\vec{x}_{\rm t}) \right] .
\label{eq:simple}
\end{equation}
Physically, $\partial \Delta z(\vec{x}_{\rm t}) / \partial x_{\rm t}$
represents an effective ramp that the top surface must climb
to escape the local potential energy minimum.
It is analogous to $\tan \alpha$ in Fig.~\ref{fig:slope},
but arises from many molecular-scale interactions.
Note that the factorization that leads to Eq. \ref{eq:simple} relies
on the exponential form of the local energy penalty.
However our numerical tests with other interaction potentials and
the simulations with Lennard-Jones potentials described below
show that the result is more general.

The effect of the area of contact on the friction coefficient $\mu_{\rm s}$
is most easily seen by making a cumulant expansion of Eq.~(\ref{eq:def_ene}).
To leading order, $\Delta z$ is proportional to the mean-squared difference
between $\delta z_{\rm t}$ and $\delta z_{\rm b}$.
Only the contribution from the cross product contributes to the variation
of $\Delta z$ with respect to $\vec{x}_t$, yielding
\begin{equation}
\label{eq:cumulant}
\mu_{\rm s} \approx
 \max_{x_t}
{1 \over \xi} \sum_{\vec{k}} i k_x
\delta \tilde{z}_{\rm b}(\vec{k}) \delta\tilde{z}_{\rm t}^*(\vec{k})
e^{i \vec{k} \cdot \vec{x}_{\rm t} } ,
\end{equation}
where
$\delta \tilde{z}_{\rm b,t}$ are the 2D Fourier transforms of
$\delta z_{\rm b,t}$.
Explicit evaluation of Eq. \ref{eq:simple} for various trial surfaces,
and the simulations described below,
show that higher order terms in $\delta z$
do not alter any of the following predictions.

Eq.~(\ref{eq:cumulant}) implies several relations between the geometry
of the two surfaces and the friction between them:
(i) If the two surfaces are crystalline and commensurate,
then $\delta z_t$ and $\delta z_b$ share a common periodicity.
The corresponding Bragg peaks in their Fourier transforms lead to
an area independent $\mu_{\rm s}$.
(ii) The magnitude of $\mu_{\rm s}$ is largest for two identical surfaces,
where all Bragg peaks contribute.
For commensurate surfaces that are not identical, $\mu_{\rm s}$
decreases exponentially with the length of the common period.
For example, if there are $p$ lattice constants of the bottom surface for
every $q$ lattice constants of the top surface, then the contribution
to $\mu_{\rm s}$ comes from the $q_{th}$ Bragg peak of the bottom surface and
the $p_{th}$ Bragg peak of the top surface.
The Fourier content of atomically rough surfaces\cite{Steele}
drops at least exponentially with $|\vec k|$,
producing a corresponding decrease in $\mu_{\rm s}$.
(iii) As a consequence of (ii), $\mu_{\rm s}$ vanishes completely for
infinite incommensurate contacts.
Contributions from the circumference of finite contacts yield
a rapidly vanishing contribution to $\mu_{\rm s}$ \cite{wenning00}.
(iv) For two disordered but smooth surfaces 
$\delta \tilde{z}_{\rm t}(\vec{k})$ and $\delta \tilde{z}_{\rm b}(\vec{k})$
have rings of diffuse scattering that overlap.
There will be many $\vec{k}$ that contribute to $\mu_{\rm s}$, but the
phase of each contribution will be random.
For interfaces with contact-area independent corrugation
$\langle \delta z_{\rm t,b}^2 \rangle$, one can immediately conclude
from the law of large numbers that
$\mu_{\rm s} \propto A^{-1/2}$.
Self-similar and curved interfaces need a more careful treatment
\cite{wenning00}.

Two geometry independent predictions can also be drawn
from Eqs.~(\ref{eq:simple}) and~(\ref{eq:cumulant}):
(v) $\mu_{\rm s}$ does not depend on the interaction strength $\varepsilon$.
(vi) Allowing for elastic deformation of the surfaces
typically reduces $\mu_{\rm s}$, because the roughness decreases
as the surfaces become more compliant \cite{footc}. 

Predictions (i) to (iii) agree with previous analytic and simulation
studies of the force between
commensurate and incommensurate
surfaces \cite{hirano90,sorensen96,muser00}.
(v) is in agreement with recent computer simulations \cite{he99},
where doubling
the strength of Lennard-Jones interactions produced almost no change
in $\mu_{\rm s}$.
(iv) and (vi) are new results that are tested by simulations below.
First we consider the implications of mobile atoms between the
surfaces within our analytic model.

\begin{figure}
\begin{center}
\leavevmode
\hbox{ \epsfxsize=50mm \epsfbox{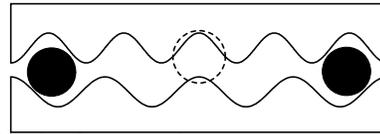} }
\caption{Sketch of two rigid, incommensurate surfaces that are separated
by a submonolayer of mobile atoms.
Circles indicate positions where the gap
between upper and lower wall is a local maximum.
The larger the gap,
the smaller the energy penalty for occupation by a mobile atom.
Full circles indicate ($+\,+$) positions
where atoms fit into concave regions of both surfaces.
The open circle indicates a less favorable
($+\,-$) site where one wall is concave and
the other is convex.
}
\label{fig:gunk}
\end{center}
\end{figure}

Fig. \ref{fig:gunk} illustrates how a mobile layer of atoms between
two surfaces can lead to a finite $\mu_{\rm s}$ {\em independent}
of the geometry of the two surfaces.
The mobile atoms are able to move to positions where they simultaneously
match the geometry of both top and bottom surfaces.
This has the effect of augmenting the height of the bottom surface
in a way that matches the undulations of the top surface, and gives an
area independent contribution to
Eq. \ref{eq:simple}.

In order to incorporate mobile
atoms into our theory quantitatively,
we presume that wherever a mobile atom sits
in the interface, the effective distance between top and bottom wall is
reduced by the diameter $d$ of the mobile atom. 
This effectively incorporates the mobile atoms into one of the walls.
The position of a
mobile atom in the interface is defined by a vector $\vec{x}_i$
so that the dimensionless areal density of the mobile atoms can be written as 
$\rho(\vec{x}) = (\pi d^2/4) \sum_i \delta(\vec{x}-\vec{x}_i)$.
For the sake of simplicity,
we assume that the mobile layer screens the direct wall-wall interaction
and that interactions between the mobile atoms can be ignored.
The latter assumption is consistent with observations
\cite{he99} that $\mu_{\rm s}$ is
insensitive to the density of mobile atoms.
The validity of both assumptions was verified through simulations
for the model system introduced below.

With this model, the total energy of the system is given by the
indirect wall-wall interaction mediated through the mobile atoms.
This takes the form
\begin{eqnarray}
\label{eq:mediated}
V_{\rm ww} & = & \varepsilon \int d^2 x \; \rho(\vec{x}) \;
  e^{-\left\{ z_{\rm t} + \delta z_{\rm t}(\vec{x}-\vec{x}_{\rm t})
- \delta z_{\rm b} (\vec{x})
   - d\right\} /\xi}\\
         & = & {{\varepsilon \pi d^2}\over{4}} {\rm e}^{-(z_{\rm t}-d)/\xi}
\sum_i e^{-(\delta z_{\rm t}(\vec{x}_i-\vec{x}_t)
-\delta z_{\rm b}(\vec{x}_i))/\xi}. \nonumber
\end{eqnarray}
The force on the top wall $\vec{F}_{\rm t}$ can again be calculated by taking
the gradient
of $V_{\rm ww}$ with respect to the center of mass position of the top wall:
$(\vec{x}_{\rm t},z_{\rm t})$. It is possible to decompose $\vec{F}_{\rm t}$
into forces $\vec{F}_{{\rm t},i}$ that stem from
individual atoms. For each such force, a linear relationship between the 
$x$ component of the force ${F}_{{\rm t},ix}$ and the ``local'' load
${F}_{{\rm t}iz}$ can be established that is similar to Eq.~(\ref{eq:simple}):
\begin{equation}
F_{{\rm t},ix} = \left[ {\partial \over \partial x_{\rm t}}
\left(\delta z_{\rm t}(\vec{x}_i-\vec{x}_t) -\delta z_{\rm b}(\vec{x}_i)
\right)
\right] F_{{\rm t},iz}.
\label{eq:sub_forces}
\end{equation}
As long as the partial derivative has a well-defined average value,
the total friction will rise linearly with the total load.

At this point, we turn back to a more qualitative discussion.
As long as the temperature is small compared to the energy barrier
for diffusion between the surfaces, atoms will sit near local energy
minima.
The simulations described below show that these local minima usually
correspond to the (++) configurations identified in Fig. \ref{fig:gunk},
where the energy with respect to both surfaces is concave.
Lateral displacements of the walls will ultimately limit the volume
accessible to (++) atoms and increase $V_{ww}$.
The atoms have created interlocking asperities that resist
sliding much like those in
Fig.~\ref{fig:slope}.
One can show more rigorously from Eq. \ref{eq:mediated}
that if all atoms sit at (++) positions, the
response to a lateral displacement is an opposing force that is linear in
the normal load $L$ and
independent of the area of contact.
Indeed, this ``Amontons' law for elastic pinning'' with an elastic
restoring force linear in $L$ has been observed 
experimentally~\cite{berthoud98}.
Thus, we may conclude that mobile atoms in the interface lead to 
Amontons' laws with a non-vanishing friction coefficient.
The pinning of two surfaces through a ``between-sorbed''
layer can also be interpreted in terms of a generalized
Frenkel-Kontorova or Tomlinson model~\cite{robbins00}.
In these models, incommensurate layers exhibit static friction when
one of the surfaces is so compliant that it can conform to the
other surface.
The mobile atoms act like an elastic sheet with nearly infinite
compliance and thus allow locking to occur.

We have tested the above predictions for disordered walls using
molecular dynamics simulations of the model described in detail
in Refs. \cite{he99} and \cite{thompson95}.
The two walls contain discrete atoms tied to their equilibrium
sites by springs of stiffness $\kappa$.
Periodic boundary conditions are applied in the plane of the walls.
Mobile molecules between the surfaces contain $n$ spherical
monomers bound into a chain.
All monomers and wall atoms also interact with a Lennard-Jones potential.
For the results presented below, $n=1$ or 6, and the energy and length scales
($\epsilon$ and $\sigma$, respectively)
of the Lennard-Jones potential are the same for monomers and wall atoms.
Other parameters produced equivalent results.
Disordered walls were made by quenching bulk fluid states and then removing
all atoms above or below some height.

\begin{figure}
\epsfxsize=70mm
\epsfbox{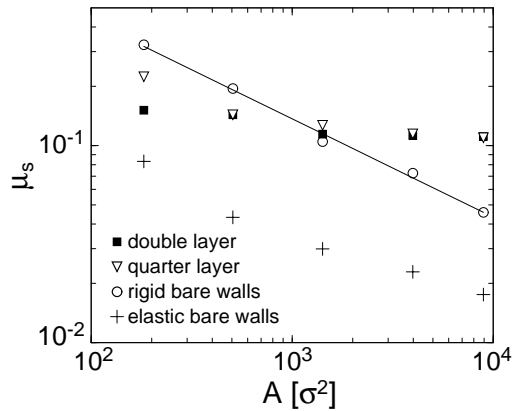}
\caption{Static friction coefficient as a function of contact area $A$.
The friction coefficient for rigid or elastic ($\kappa=400\epsilon \sigma^{-2}$)
bare walls drops as $A^{-1/2}$ (solid line).
Inserting enough mobile molecules ($n=6$) to make either
a quarter or two monolayers
yields nearly the same area independent value of $\mu_s$.
Statistical error bars are comparable to the symbol size.
}
\label{fig:wwogunk}
\end{figure}

Fig. \ref{fig:wwogunk} shows representative results for the scaling
of friction with contact area.
The top wall was coupled to a constant normal load and a slowly increasing
lateral force.
The static friction was determined from the threshold force needed
to initiate lateral motion of the top wall.
For all cases considered, the coefficient of friction between bare disordered
surfaces vanished as $A^{-1/2}$ (solid line)
in agreement with prediction (iv).
In contrast, when molecules were inserted into the interface,
the static friction rapidly approached a constant as $A$ was increased.
It is interesting to note that the bare and disordered surfaces have
equal friction at an area corresponding to about 1000 atoms.
This is slightly larger than contacts in atomic force microscopy,
but much smaller than the micron size contacts between typical surfaces.
In agreement with prediction (vi),
decreasing $\kappa$ to the smallest reasonable value \cite{muser00}
($\kappa=400\,\epsilon \sigma^{-2}$) reduced the friction between bare
walls considerably (crosses).
Much smaller reductions
($\sim$20\%) were observed when mobile atoms were present between elastic
surfaces.
In both cases the
reductions reflect the ability for atoms to move vertically to
minimize the steric overlap with the opposing surface.
When atomic displacements were constrained to the horizontal plane,
we observed no reduction in $F_{\rm s}$.

\begin{figure}
\epsfxsize=70mm
\epsfbox{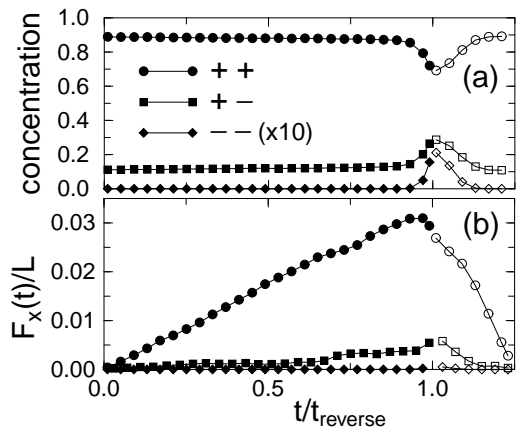}
\caption{(a) Fraction of atoms in (+\,+), (+\,-), and (-\,-) sites (Fig. 2)
and (b) their
contribution to the restoring friction force as a function of time.
For these runs $t_{\rm reverse}$ was about 300 in Lennard-Jones units.
}
\label{fig:conc}
\end{figure}

The simulations also allowed us to test the assumption that monomers in
(++) configurations (Fig. \ref{fig:gunk}) dominate the pinning.
Fig. \ref{fig:conc} shows results for a quarter monolayer of spherical
molecules ($n=1$) as the force was ramped up to a value slightly
above $F_{\rm s}$ over a time $t_{\rm reverse}$
and then returned four times as rapidly to zero.
About 90\% of mobile atoms sit at (+\,+) positions until the wall begins
to slide ($t / t_{\rm reverse} \approx 0.95$),
and they contribute an even larger fraction of $F_{\rm s}$.
Atoms that sit at convex positions relative to both walls (-\,-),
were only seen during sliding.
Even in this dynamic state, almost 70\% of the atoms are in (+\,+) sites
and they continue to provide most of the lateral force.
A detailed
analysis of these runs shows
that the mechanisms of kinetic and
static friction are closely related in this model.
At any instant in time, most of the mobile atoms are in local energy
minima and resist lateral motion.
When these sites become unstable (+\,-) or (-\,-) sites, the
atoms pop rapidly to a new site.
Energy is dissipated during these rapid pops and flows into the
walls as heat.

In conclusion, we have presented a simple analytic model for the molecular
origins of friction.
Although the model assumes exponential repulsion between surfaces,
we presented
simulations with more realistic potentials that show that the predictions
are more generally applicable.
Specifically the model provides a microscopic foundation for
Amontons' law $F_{\rm lateral} = \mu L$,
but shows that $\mu$ vanishes for bare incommensurate or disordered
surfaces as the size of contacts increases.
Introducing mobile atoms between the surfaces yields Amontons' laws:
$\mu$ is independent of surface area and load for any contact geometry.
The simulations also reveal deep connections between static and
kinetic friction.
Of course, experimental systems contain additional features that
have not been treated.
These include long-range elastic and plastic deformations of the walls,
generation of wear debris, and chemical reactions \cite{bowden86}.
All of these may be important in particular systems.

We thank K. Binder, G. He and B. N. J. Persson for useful discussions.
Support from the National Science Foundation through Grant No. DMR-9634131
is gratefully acknowledged. MHM is grateful for support through
the Israeli-German D.I.P.-Project No 352-101.

\end{multicols}

\begin{thebibliography}{99}
\bibitem{footc}
This trend is reversed in the limit were the solids become
extremely soft, allowing them to deform into locally commensurate
regions \cite{hirano90,muser00}.
Detailed analysis shows that this limit corresponds to having
stronger interactions between atoms on opposing surfaces than
within each surface \cite{muser00,robbins00}.
\end{thebibliography}
\end{document}